\documentclass[aps,pre,longbibliography,twocolumn,floatfix]{revtex4-1}
\usepackage{graphicx}

\usepackage{commath}
\usepackage{amsmath}
\usepackage{amssymb}
\usepackage{bm}

\usepackage{mathtools}
\usepackage{times}
\usepackage[pdftex]{hyperref}
\hypersetup{colorlinks=true,linkcolor=blue,citecolor=blue,urlcolor=blue}
\usepackage{verbatim}

\begin{document}

\title{Marginally jammed states of hard disks in a one-dimensional channel}

\author{Yuxiao Zhang}
\author{M.~J.~Godfrey}
\author{M.~A.~Moore}
\affiliation{Department of Physics and Astronomy, University of Manchester,
Manchester M13 9PL, United Kingdom}

\date{\today}

\begin{abstract}
  We have studied a class of marginally jammed states in a system of
  hard disks confined in a narrow channel---a quasi-one-dimensional
  system---whose exponents are not those predicted by theories valid
  in the infinite dimensional limit.  The exponent $\gamma$ which
  describes the distribution of small gaps takes the value $1$ rather
  than the infinite dimensional value $0.41269\dots$.  Our work shows
  that there exist jammed states not found within the tiling approach
  of Ashwin and Bowles.  The most dense of these marginal states is an
  unusual state of matter that is \emph{asymptotically} crystalline.
\end{abstract}
\maketitle

\section{Introduction}
\label{intro}
There has been remarkable progress in our understanding of the
properties of marginally \cite{wyart:12,muller:15} jammed states of
hard spheres \cite{Altieri:2019,charbonneau:12,sevelev:17, Kurchan:13,
  Charbonneau:14,Charbonneau:17}.  The most striking predictions are
for the values of the exponents that describe the geometry of the
marginally jammed system of hard spheres, such as the distribution of
small gaps $h$ between the spheres, $ g(h) \sim 1/ h^{\gamma}$, as $h
\to 0$.  The exponent $\gamma=0.41269\dots$, according to the
infinite-dimensional replica-symmetry-breaking \cite{parisi:79} mean
field theory \cite{Charbonneau:14}.  Numerical results in dimensions
$d = 3$ to~$10$ \cite{charbonneau:12} are consistent with the exponent
being independent of dimension and equal to its value for
$d\to\infty$.

This is a striking result which remains unexplained.  Indeed, possible
issues with the infinite-dimensional approach applied to $d=3$ have
been noted in \cite{procaccia:18}.  There is also evidence that the
exponents might not actually be independent of dimension, at least for
shear-driven (and hence anisotropic) jammed states \cite{olsson:19}.
Despite this, it is commonly expected that the ``upper critical
dimension'' is $2$ \cite{wyart:05,Wyart:06, Liu:10,
  Goodrich:12,charbonneau:12,hexner:18,hexner:19}.  That is, the exponents should maintain
their infinite-dimension values down to two dimensions; this
expectation is supported by numerical evidence for logarithmic
corrections to scaling in $d = 2$
\cite{goodrich:14,hecke:14,vagberg:16}.  In this paper we present
evidence that marginally-jammed states in a one-dimensional system of
hard disks have very different values for the exponents; for example,
we find that $\gamma=1$ exactly.  Our result therefore provides
further evidence for an upper critical dimension for marginally-jammed
packings. The arguments of Refs. \cite{wyart:05, Wyart:06,hexner:18,hexner:19} are Imry-Ma style arguments for $2$ as the upper critical dimension: they do not provide numerical values for (or even demonstrate the existence of) exponents such as $\gamma$ if the system dimension is below 2. The present work does both.

The one-dimensional system of hard disks in a narrow channel has been
extensively studied in other contexts
\cite{Robinson:16,Godfrey:15,Godfrey:14,bowles:06,Yamchi:12,Ashwin:13,
  Yamchi:15, Kofke:93, Varga:11,
  Gurin:13,charbonneau:18d,hicks:18,Huerta:20}.  The channel width
available to the centers of the disks has in this paper the value
$\tilde{h}=0.95 \sigma$ so that the disks, which have diameter $\sigma$,
cannot pass each other, but can touch their nearest and next-nearest
neighbors, as shown in Fig.~\ref{fig:y_coords}.  The packing fraction
$\phi$ is defined as $\phi=N\pi\sigma^2/(4 H_d L)$, where $N$ is the
number of disks in a channel of length $L$ and width $H_d=\tilde{h}+\sigma$.

The jammed states for channel widths $\tilde{h} < \sqrt{3}\sigma/2$ have
already been studied \cite{Ashwin:13, godfrey:18}.  The complexity of
these jammed states (which is the logarithm of the total number of jammed states divided by $N$) is non-zero and all of the states are isostatic, in
that there are just enough contacts to ensure stability.
However, when $\tilde{h}<\sqrt{3}\sigma/2$ there are no small gaps at all
between the disks, whereas for the range of $\tilde{h}$ studied in this paper,
$\sqrt{3}\sigma/2 <\tilde{h} <\sigma$, there are gaps that can be arbitrarily
small: their distribution is described by the exponent $\gamma = 1$.
The jammed states with arbitrarily small gaps have non-zero complexity
and are isostatic; however, there are also jammed states in this system that are not isostatic but hyperstatic, and these have a much higher complexity according to the results of Ashwin and Bowles \cite{ashwin:09}.

Recently Ikeda \cite{ikeda:20} has studied disks in a narrow channel when the diameters
of the disks have a dispersion in size and obtained $\gamma =0$ for the gap exponent. Polydispersity like this is often used in numerical simulations of glasses  to speed up the simulations \cite{berthier:19c}, or to discourage crystallization. Polydispersity is unphysical for molecular glasses: molecules actually have the same size. For disks in a narrow channel crystallization does not occur, except at the highest possible density. A small amount of polydispersity will convert the hyperstatic jammed states of the type discussed by Ashwin and Bowles into isostatic states, just as happens for crystals \cite{mari:09, charbonneau:18}.  Ikeda's work allows a study of whether its use is harmless by finding the value of $\gamma$ for the isostatic states which it produces. Ikeda \cite{ikeda2019} has  argued that above the upper critical dimension, exponents like $\gamma$ should not depend on whether polydispersity has been used to produce the isostatic states. This claim is apparently at variance with the simulational study of Ref.~\cite{charbonneau:18} in $d=3$, which is above the upper critical dimension. In $d=1$ (below the upper critical dimension),  our isostatic states  have  $\gamma = 1$, which is very different from the case of isostatic states obtained by adding polydispersity: these have $\gamma = 0$, according to Ref.~\cite{ikeda:20}. 

Ashwin and Bowles \cite{ashwin:09} had in fact claimed to have
constructed the complete jamming landscape for disks all of the same
size when $\sqrt{3}\sigma/2 <\tilde{h} <\sigma$.  They argued that all
the jammed states in the system could be constructed from 32 varieties
of tile together with rules limiting which tiles could be joined to
other tiles.  Figure~\ref{fig:y_coords} shows two configurations that
can be described by their tiling procedure: a unit cell of the state
of maximum density (the ``buckled crystal") and an asymmetric
pentagon.  In the tiling picture, the gaps between disks take on a
finite number of nonzero discrete values.  However, we have discovered
that the Ashwin-Bowles picture is incomplete: there are jammed states
that would require an \emph{infinite} number of tiles.  We first
became aware of this by studying the jammed states produced by
Lubachevsky-Stillinger \cite{Lubachevsky:1990} quenches: these states
regularly contained jammed disks at distances from the walls that were
not among those considered by Ashwin and Bowles. In fact, some
quenched states include short ``sandwich states"; for example, in the
jammed state shown in Fig.~\ref{fig:towards_teal}(g), the region to
the left can be constructed from the Ashwin-Bowles tiles, while the
right-hand portion consists of a short sandwich state.  As a
consequence of this, we shall not be using the tiling approach.

Our work is mainly analytical; we resort to numerical methods only to
verify the stability of the jammed states. The numerical methods are
described in the Appendix. In Sec.~\ref{algo} we describe the
algorithm used to generate the sandwich states. In
Sec.~\ref{stability} we show that these states are stable in some
ranges of $N$. In Sec.~\ref{gamma} we determine the value of the
exponent $\gamma$ in these jammed states and investigate the
distribution of small contact forces to determine another commonly
studied exponent, $\theta$.

\section{The algorithm}
\label{algo} 

\begin{figure}
  \includegraphics[width=\columnwidth]{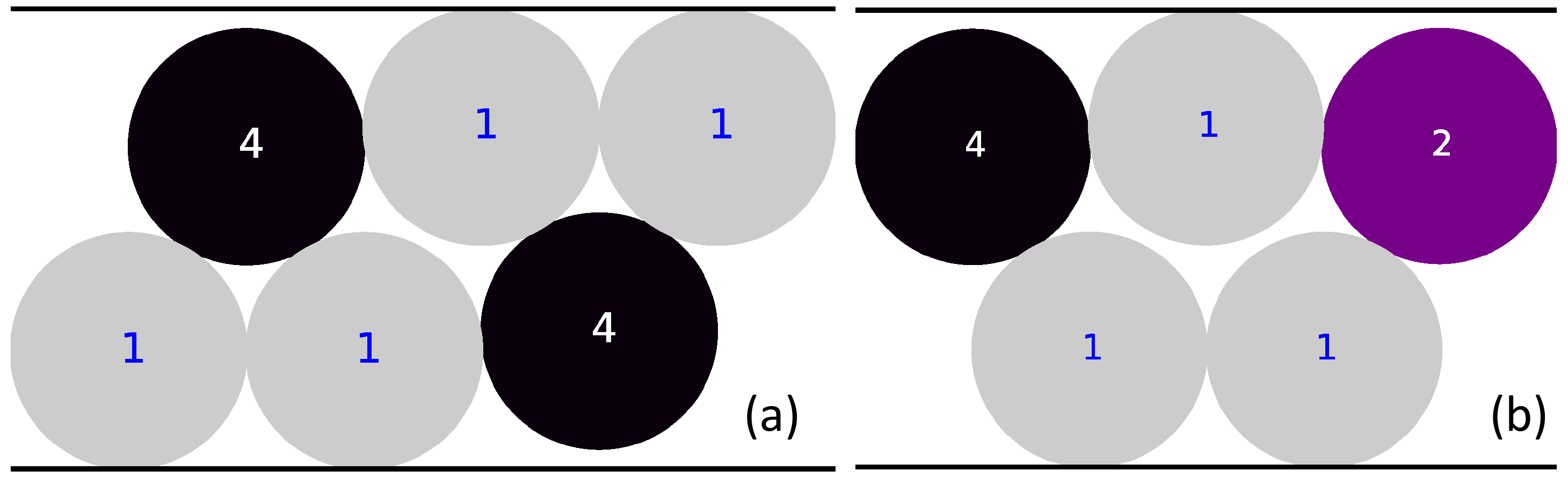}
  \caption{(a) A unit cell of the buckled crystal, which is the jammed
    state of maximum density.  Its packing fraction is $\phi \simeq
    0.8074$.  (b) An asymmetric pentagon, which is one of the
    Ashwin-Bowles tiles \cite{ashwin:09}.  Labels $1,2,4$ refer to
    particular values of $y$, as described in the text.  The color
    coding is given in Fig.~\ref{fig:towards_teal}.}
  \label{fig:y_coords}
\end{figure}

\begin{figure}
  \includegraphics[width=\columnwidth]{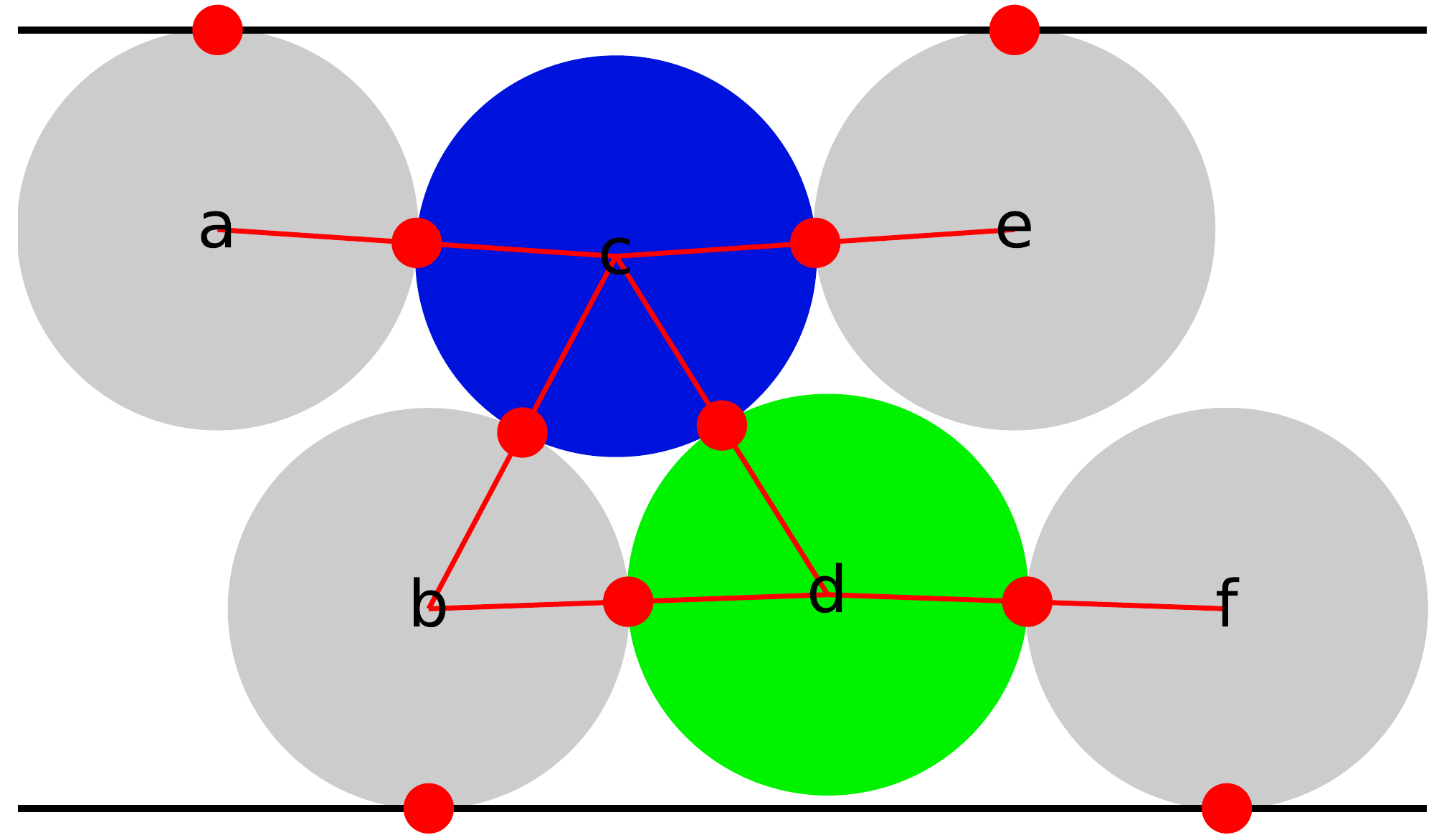}
  \caption{Diagram to illustrate an algorithm for generating jammed
    states, as discussed in the text.  Red dots denote contacts and
    red lines join the centers of disks that are in contact.  The
    color coding follows Fig.~\ref{fig:towards_teal}.}
  \label{fig:algo}
\end{figure}

\begin{figure*}
  \includegraphics[width=1.0\textwidth]{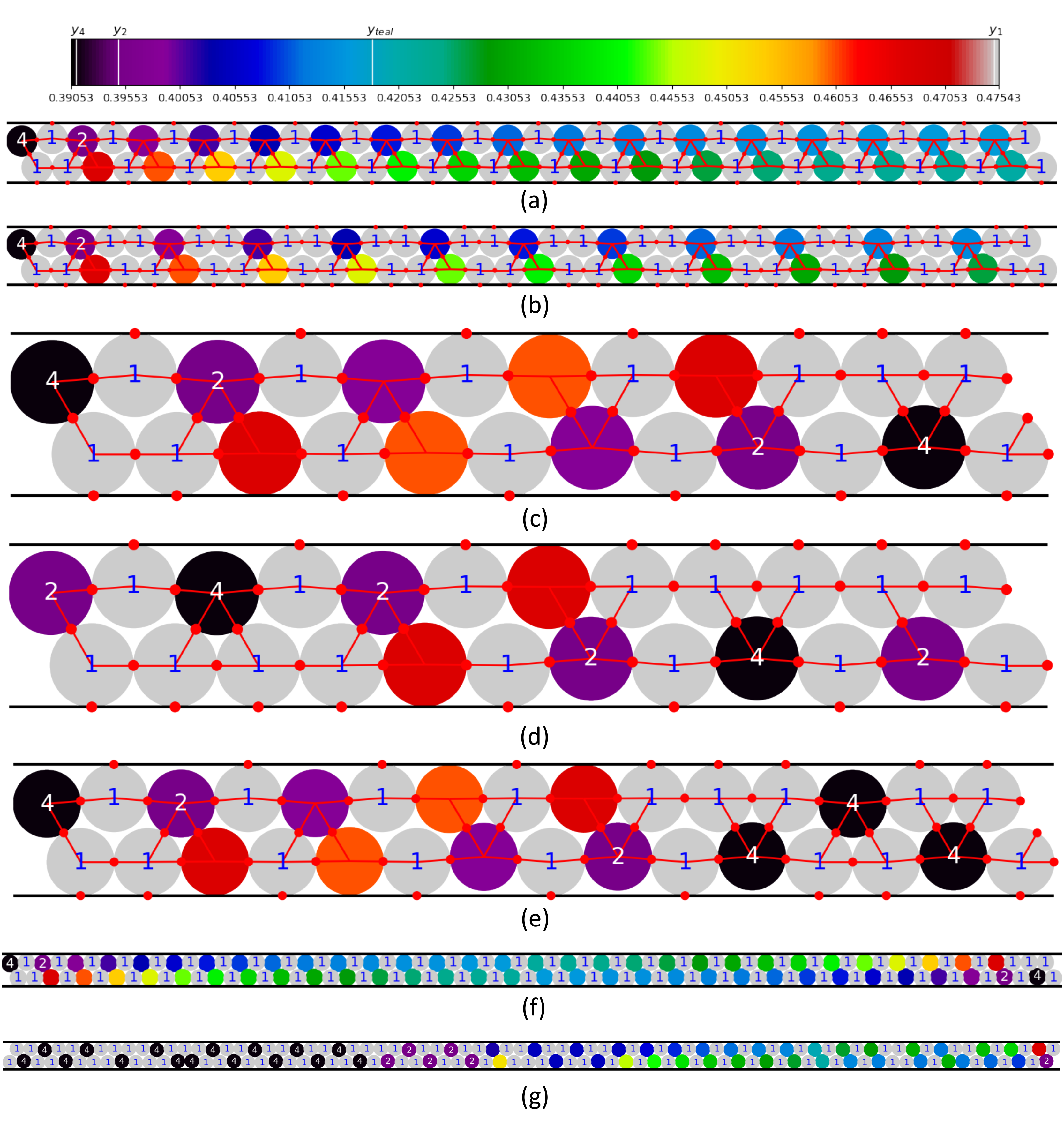}
  \caption{Top: Color code for the $y$-coordinates of disks.  (a)
    Colored disk pairs progressively approach the y-coordinates $\pm
    y_{\rm teal}=\pm0.41808$ associated with the color teal.  (b) A
    jammed state where two pairs of disks have been inserted in
    positions $y_1$.  (c) A periodic ``sandwich'' state for $N=24$.
    It has packing fraction $\phi=0.80686$.  This state is only
    locally jammed: on compression it evolves to the state (d), whose
    packing fraction is $\phi=0.80688$.  (e) A system of $N=30
    =(24+6)$ disks containing \textit{two} unit cells of buckled
    crystal, which produces a state with more contacts.  It is stable
    and has packing fraction $\phi=0.80697$.  (f) A stable sandwich
    state with $N = 128$. (g) A jammed state for $N=150$ obtained after a Lubachevsky-Stillinger \cite{Lubachevsky:1990} quench from an initial state of packing fraction $\phi=0.70$ to a final packing fraction $\phi=0.807169$. }
  \label{fig:towards_teal}
\end{figure*}

\begin{figure}
  \includegraphics[width=\columnwidth]{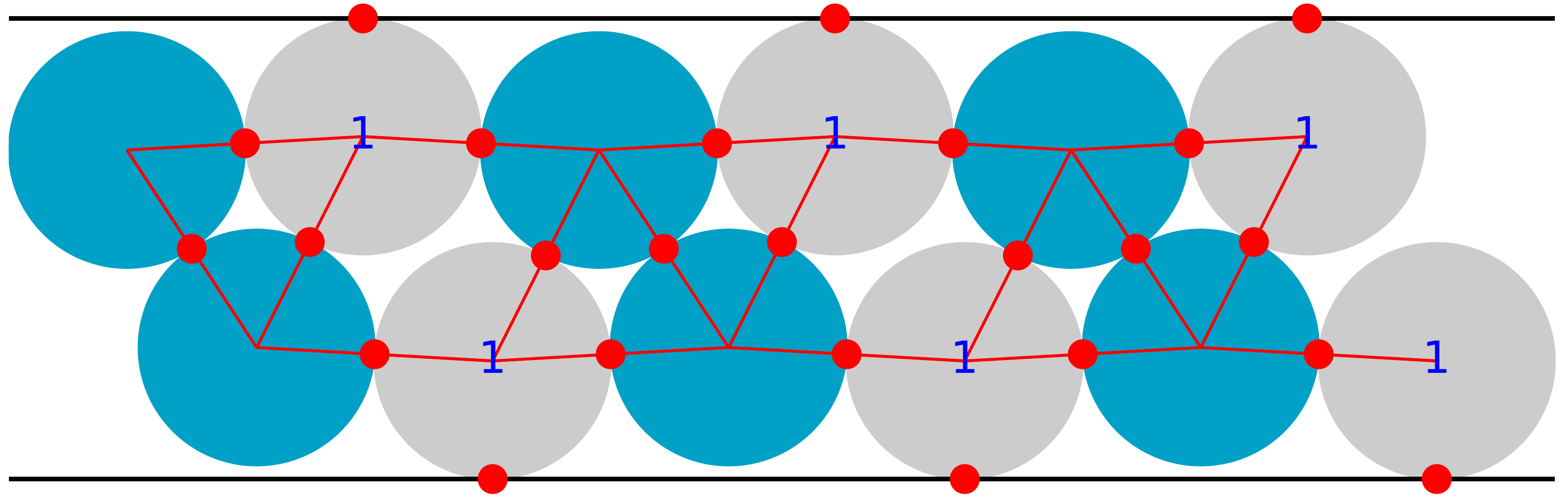}
  \caption {Three unit cells of the limiting state that is approached
    in Fig.~\ref{fig:towards_teal}(a).  The packing fraction $\phi
    \simeq 0.8068$.  Note that in Fig.~\ref{fig:towards_teal}(a) there
    is always a gap between each pair of disks corresponding to disks
    (d) and (e) in Fig.~\ref{fig:algo}; this gap is zero for the
    limiting state.}
  \label{fig:teal}
\end{figure}

The coordinates of a disk will be labelled $(x,y)$, where $y$ is
measured from the mid-point of the channel.  A disk which is just
touching a wall has $y=\pm y_1$, where $y_1=\tilde{h}/2$; it will be referred
to as a $1$-disk (see Fig.~\ref{fig:y_coords}).  A $4$-disk may have
positions $\pm y_4$, where $y_4=\sqrt{3} \sigma/2-\tilde{h}/2\simeq0.391025$ (we set $\sigma=1$ in this paper).
It is at the apex of an equilateral triangle whose other vertices are
$1$-disks.  One further disk position, $y_2$, frequently arises in
quenches; it is defined by the pentagonal arrangement of disks shown
in Fig.~\ref{fig:y_coords}(b).

The jammed states that we focus on in this paper (see
Fig.~\ref{fig:towards_teal}) consist of tilted equilateral triangles
of disks, separated by at least one pair of $1$-disks.  Refer to
Fig.~\ref{fig:algo} which shows six disks (a,b,c,d,e,f) and their
contacts with each other and the sides of the channel: (a) and (b) are
nearest-neighbors lexically, and (a) and (c) next-nearest neighbors
lexically (but (c) is physically closer to (a) than (b)).  The
triangles have varying orientations that are obtained via the
following algorithm:
\begin{enumerate}
  \item Disk (a) touches the wall, $y_a=\pm y_1$.  An equilateral
    triangle of disks (b,c,d) has its first disk, (b), touching the
    opposite wall, $y_b=\mp y_1$, and its second disk, (c) (the apex of the equilateral triangle), touching
    disk~(a).  Given the values of $(x_a,y_a)$ and $(x_b,y_b)$, the
    positions of disks (c) and (d) can be obtained exactly, e.g. by
    solving the four equations specifying the contacts (ac), (bc),
    (bd), and~(cd).
  \item Two more disks, (e) and (f), are placed at $\pm y_1$, touching
    their next-nearest neighbors, (c) and (d), respectively.
  \item Steps 1 and 2 are repeated with disks (e) and (f) replacing
    (a) and~(b), respectively.
\end{enumerate}
At step (2) one could, instead of placing one pair of disks, choose to
put in two pairs, as shown in Fig.~\ref{fig:towards_teal}(b): the
stability of the jammed state is unaffected.  States with a nonzero
complexity can be obtained by placing the additional pairs of disks at
random in the configuration generated by the deterministic algorithm
1--3.

The algorithm 1--3 generates configurations like those shown in
Fig.~\ref{fig:towards_teal}(a) and (b).  The $y$ coordinates of the
colored disks approach a limiting value, $y_{\rm teal}$, which in our
diagrams is represented by the color teal (intermediate between green
and blue).  The limiting state is crystalline; it is depicted in
Fig.~\ref{fig:teal}.

The jammed states in Fig.~\ref{fig:towards_teal}(a) and (b) do not
satisfy periodic boundary conditions.  Periodic states can be
constructed as shown in Fig.~\ref{fig:towards_teal} (c), (e) and (f):
two regions containing tilted equilateral triangles of disks are
sandwiched between disks at positions 1 and~4, which form a region of
buckled crystal.  These sandwich states are characterized by
an inversion of the tilted equilateral triangles near their midpoint.

\section{Marginal Stability}
\label{stability}
We next discuss the stability of the sandwich states.  They are
manifestly \textit{locally} jammed \cite{donev:04}, as this only
requires each disk to have three contacts, not all of which lie on the
same half of the disk's perimeter. We have checked these states for
stability against a collective displacement of any number of disks by
using an extension of the linear programming algorithm discussed in
Ref.~\cite{donev:04}; details can be found in the Appendix. The algorithm not only checks for stability but
also finds a stable state that the system can reach when it is not
stable, as illustrated in Fig.~\ref{fig:towards_teal}(d) for the case
$N= 24$.  (Sandwich states with just one unit cell of buckled crystal
within them have values of $N$ which are multiples of 8.)  We find that the sandwich states containing one
unit cell of buckled crystal are stable for $N =8$ and $N =16$ and for
all $N \ge 96$.  However, for $N =24$, $32$, $\dots, 88$, the sandwich
states can be made stable by the addition of a second unit cell of
buckled crystal; an example is shown in Fig.~\ref{fig:towards_teal}(e)
for the case $N = 24 +6 =30$.

The sandwich states can be regarded as marginally stable for
large~$N$.  A truly marginal state is one in which the number of
contacts of the disks is equal to the number of their degrees of
freedom $N_f$.  The latter is $2N-1$ (not counting the uniform
translation of the entire system along the $x$-axis); the number of
contacts in periodic sandwich states with one unit cell of buckled
crystal as shown in Fig.~\ref{fig:towards_teal}(c) is $N_c=2N+2$, so
$N_c-N_f=3$; i.e., there are just three more constraints than required
for strict marginality.  Thus, for large $N \,(\ge 96)$, the sandwich
states approach the condition of marginal stability, in the sense that
the ratio $(N_c-N_f)/N$ can be made arbitrarily small by choosing a
sufficiently large value for~$N$.

\begin{figure}
  \includegraphics[width=\columnwidth]{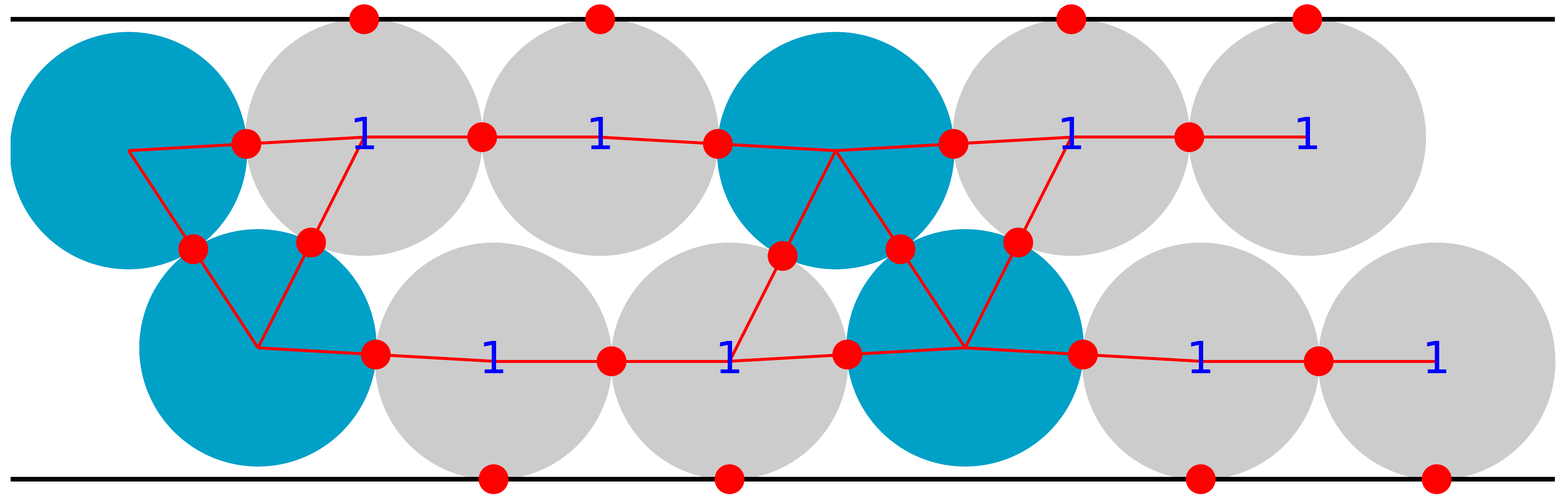}
  \caption{Two unit cells of the limiting state that contains the
    maximum number of 1--1 insertions.  It has packing fraction $\phi
    \simeq 0.8064$.}
  \label{fig:11inserts}
\end{figure}

\section{Determination of the exponent $\gamma$}
\label{gamma}
The marginally stable configurations have gaps that become vanishingly
small for $N\to\infty$.  They are the gaps between disks of types (d)
and (e) in Fig.~\ref{fig:algo}; see also
Fig.~\ref{fig:towards_teal}(a).  From the geometry of the state, the
sizes $h_i$ and $h_{i+1}$ of consecutive gaps can be related
algebraically.  For large values of $i$, the relation simplifies to
\begin{equation}
  h_{i+1}=h_i/b,
  \label{eq:decrease}
\end{equation}
where the constant $b \simeq 1.219$ for $h=0.95\sigma$.  This
implies that for an infinite sequence the distribution of gaps would
be of the form
\begin{equation}
  g(h)=\frac{1}{h \ln b },
\label{eq:phdist}
\end{equation}
as $h \to 0$; thus, the exponent for gap sizes is $\gamma =1$.  A
distribution varying as $1/h$ at small $h$ is not normalizable, but
this is not a problem in practice because, for any finite value of
$N$, there is a smallest gap whose size is of order $\exp(-cN)$, where
$c>0$.  The result $\gamma=1$ should be contrasted with
$\gamma=0.41269 \dots$ for jammed states of spheres in the limit
$d\to\infty$; the latter value of $\gamma$ is also consistent with
simulation results for all $d\ge3$ \cite{charbonneau:12}.

\begin{figure}
  \includegraphics[width=\columnwidth]{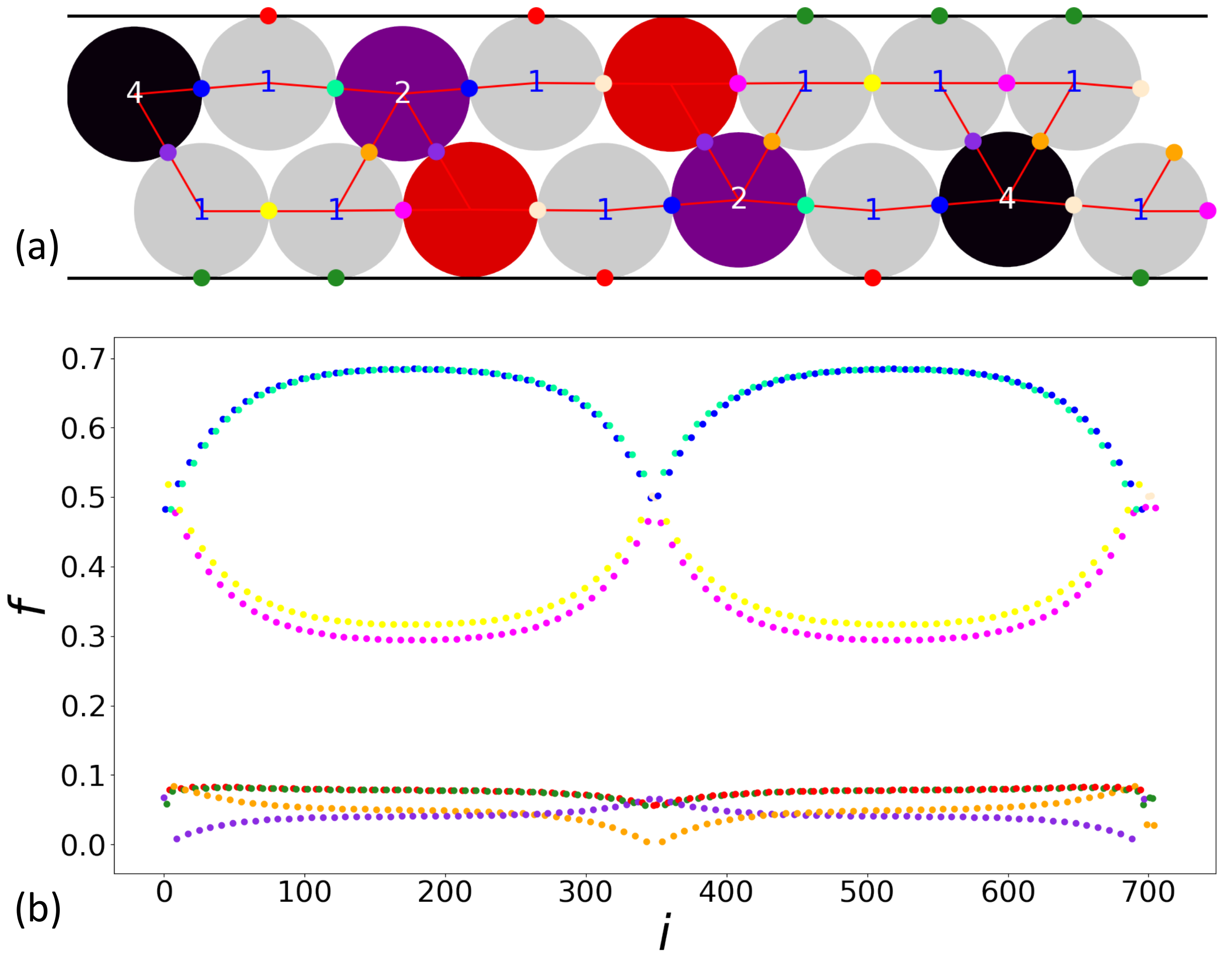}
  \caption{(a) Color coding of contacts.  (b) The force $f$ at the
    $i$th contact in a sandwich state with $N = 352$ and $N_c=706$,
    color coded as in~(a).  The smallest force is associated with the
    orange contacts at the center of the system, where the orientation
    of the equilateral triangles switches.}
  \label{fig:force}
\end{figure}

Extra 1--1 disk insertions do not change the form of $g(h)$.  Neither
do they change the value of $N_c-N_f$; it remains at~$3$.  Because any
of the 1--1 pairs can be replaced by two 1--1 pairs, the total number
of possible insertions is $2^{N/4-3}$, where $N$ is the number of disks
before the insertions are made.  When $N$ is large, the resulting
states have packing fractions in the range $0.8064 < \phi < 0.8068$;
the jammed states at the upper and lower limits of the density range
are shown in Figs~\ref{fig:teal} and~\ref{fig:11inserts},
respectively.  Within this range of densities, the sandwich states are
marginal and have a nonzero complexity.

States of the type discussed above consist of short lengths of buckled
crystal alternating with noncrystalline regions of fixed length.  Our
construction can clearly be generalized to create aperiodic structures
in which the lengths of the buckled-crystal and noncrystalline regions
vary.  We note that the distribution of small gaps, $g(h)$, will in
general be related to the distribution of lengths of the
noncrystalline regions, because the size of the smallest gap in one of
these regions depends (exponentially) on its length.  It may also
be noted that each additional region of buckled crystal increases the
excess number of contacts, $N_c-N_f$, and so takes the system further
from marginality.

Another much studied quantity for marginal jammed states is the
exponent associated with the distribution of small contact forces $f$,
$P(f)\sim f^{\theta}$.  We used the algorithms in
Ref.~\cite{charbonneau:15c} to determine the forces in the contacts
when a unit force is applied along the $x$ axis.  The smallest force,
which occurs near the center of the sandwich state (see
Fig.~\ref{fig:force} for the case $N=352$), seems to approach a
nonzero value ($\approx 0.005$) as $N \to \infty$.  The existence of a
gap in $P(f)$ suggests that $\theta = \infty$.  Notice that the
inequality \cite{lerner:13} $\gamma \ge 1/(2+\theta)$ is satisfied by
these values, as is the inequality $\gamma \ge (1-\theta)/2$.

\section{Discussion}
\label{discussion}
The sandwich state of Fig.~\ref{fig:towards_teal}(f) is a curious
state of matter.  In the limit of large $N$ it approaches the
crystalline state of Fig.~\ref{fig:teal}.  However, it never quite
reaches that state: the equilateral triangles of disks all have
slightly different orientations and the small gaps persist.  If one
were to compute the structure factor \cite{hof:95} of the state it
would have for $N\to\infty$ the delta function peaks of the crystal in
Fig.~\ref{fig:teal}.  It could thus be regarded as an asymptotic
crystal.  But if one regards it as a crystal, its unit cell is the
length of the system.  We suspect that similar states might already
have been found in other systems with restricted geometries; for
example, the packing of spheres within a long cylinder, as considered
in Ref.~\cite{charbonneau:16x}, where certain jammed states,
constructed so as to satisfy periodic boundary conditions, were found
to have a unit cell that was as large as the system.

The study of Ikeda \cite{ikeda:20} of the jammed states of disks in a narrow channel  in the presence of polydispersity produced a value of $\gamma = 0$, in contrast to our value in the absence of polydispersity, $\gamma=1$. It is commonly assumed \citep{ikeda2019} that introducing polydispersity is ``harmless"; it merely serves to allow a speed up of simulations or to prevent crystallization. Our work shows that sometimes it completely changes the physics of the problem.

\begin{acknowledgments}
  We should like to thank P. Charbonneau, and E. Corwin for
  discussions.  We would especially like to thank Harukuni Ikeda for
  many exchanges.
\end{acknowledgments}

\appendix*
\section{The length-compression algorithm}

This Appendix describes how we used a linear programming
method to determine the stability of jammed states of hard disks in a
channel.  Although the results are specific to our system, the methods
are general and have been derived and discussed in detail by A. Donev
and coworkers~\cite{donev:04}.  Our discussion of the methods is
therefore brief, and the reader is referred to Ref.~\cite{donev:04}
for a careful description.

\begin{figure}
  \centering
  \includegraphics[width=0.45\textwidth]{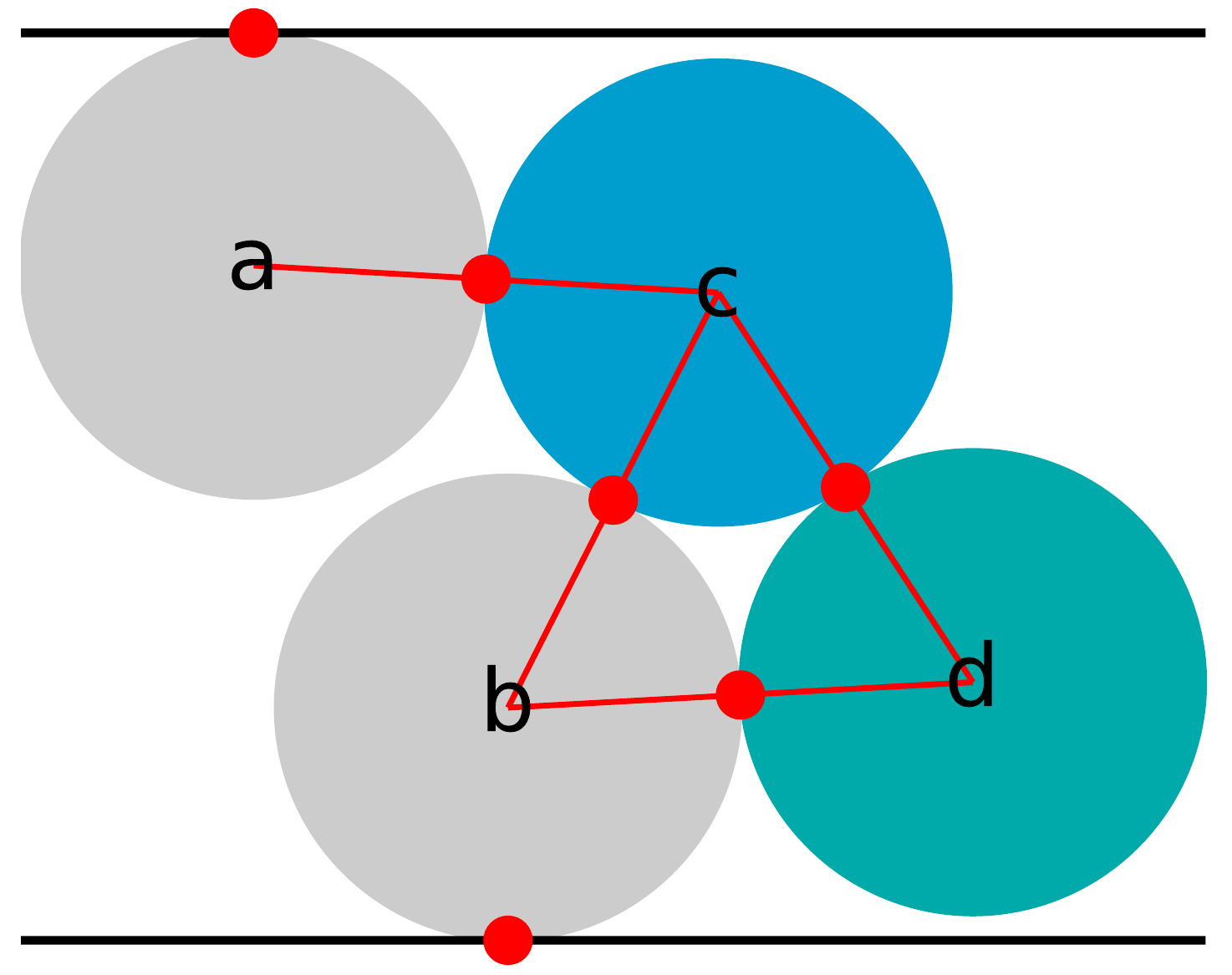}
  \caption{A configuration of four disks, used in the example of a
    rigidity matrix given in Eq.~(\ref{eq:rigidity}).  In addition to
    the actual contacts shown by red dots, the rigidity matrix
    includes potential contacts, such as those between disks (a) and
    (b), and between disks (c) and (d) and the walls.}
  \label{fig:alg}
\end{figure}

If a jammed configuration of hard disks is unstable, that is, if it
allows a collective motion of an arbitrary number of disks, then it
can be compressed until the system is truly jammed.  In our work we
consider periodic configurations of $N$ disks in a channel and
minimize their length, subject to the constraints of no overlap of
disks with each other or the channel walls.  The length of the
configuration, $L=x_{N+1}-x_1$, is a linear function of the
$2N$-dimensional vector of disk coordinates, $\mathbf{R}$.  Thus, it
is natural to consider a linear programming problem in which the
function $L[\mathbf{R}]$ is minimized, subject to the following
linearized constraints
\begin{align*}
  &\boldsymbol{\mathsf{A}}^T \Delta \mathbf{R} \geq -\mathbf{h},\\
  &\Delta x_{N+1} \leq 0,\\
  &\Delta x_1 = 0,\\
  &\Delta \mathbf{r}_{N+1} = \Delta \mathbf{r}_1,\\
  &\Delta \mathbf{r}_{N+2} = \Delta \mathbf{r}_2,\\
\end{align*}
where $\Delta \mathbf{r}_n = (\Delta x_n,\Delta y_n)^T$, $\Delta
\mathbf{R} = (\Delta \mathbf{r}_1,\Delta \mathbf{r}_2,...,\Delta
\mathbf{r}_N)^T$ is a set of potential unjamming displacements,
and $\boldsymbol{\mathsf{A}}$ is the \textit{rigidity matrix}
containing all potential contacts (i.e., pairwise contacts with all
nearest and next-nearest neighbours).  The components of the vector
$\mathbf{h}$ are the sizes of the gaps at each of the potential
contacts; the gap sizes are nonnegative.  The matrix
$\boldsymbol{\mathsf{A}}$ has two rows for each disk and $N_C$ columns
that represent the $N_C$ possible contacts.  For the example shown in
Fig.~\ref{fig:alg}, $\boldsymbol{\mathsf{A}}$ is given by

\begin{widetext}
\begin{figure*}
  \centering
  \includegraphics[width=1\textwidth]{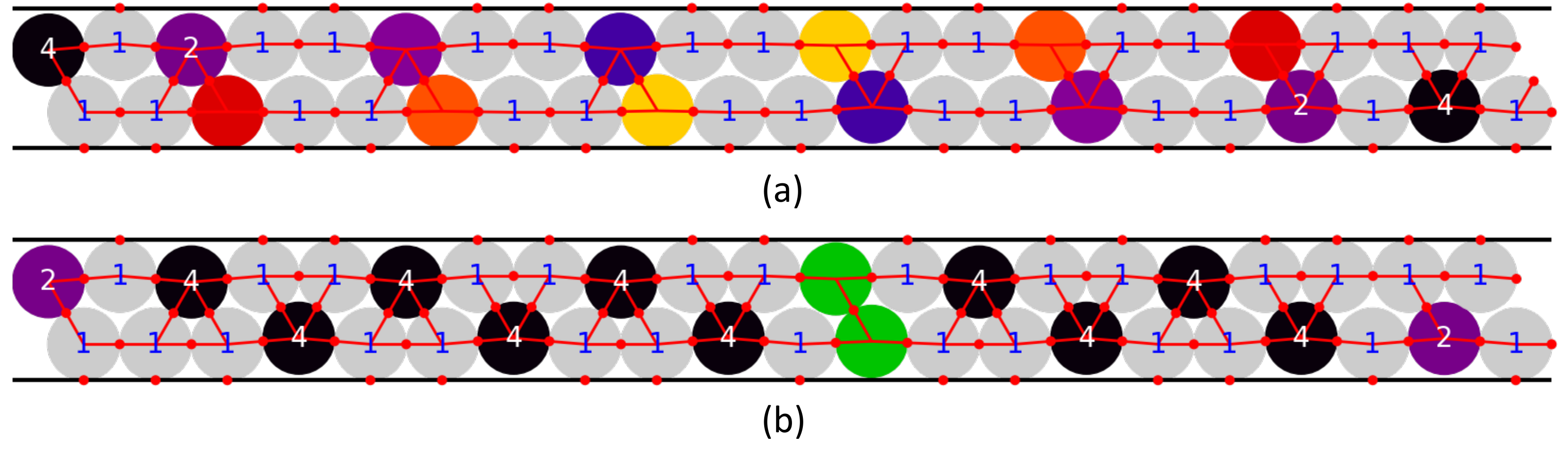}
  \caption{Before, (a), and after, (b), the compression of an unstable
    configuration of $N=42$ disks that contains the maximum possible
    number of 1--1 insertions in a sandwich state of $32$ disks.  The
    packing fraction of state~(a) is $\phi=0.80652$, which increases
    to $0.80721$ on compression to the state~(b).}
  \label{fig:compressed_w_insertion}
\end{figure*}

\begin{equation}
\boldsymbol{\mathsf{A}}
 = \bordermatrix{~ & a,w & a,b & a,c & b,w & b,c & b,d & c,w & c,d & d,w \cr
	      a & \mathbf{u}_{a,w} & \mathbf{u}_{a,b} & \mathbf{u}_{a,c} & 0 & 0 & 0 & 0 & 0 & 0 \cr
	      b & 0 & -\mathbf{u}_{a,b} & 0 & \mathbf{u}_{b,w} & \mathbf{u}_{b,c} & \mathbf{u}_{b,d} & 0 & 0 & 0 \cr
	      c & 0 & 0 & -\mathbf{u}_{a,c} & 0 & -\mathbf{u}_{b,c} & 0 & \mathbf{u}_{c,w} & \mathbf{u}_{c,d} & 0 \cr
	      d & 0 & 0 & 0 & 0 & 0 & -\mathbf{u}_{b,d} & 0 & -\mathbf{u}_{c,d} & \mathbf{u}_{d,w} \cr},
\label{eq:rigidity}
\end{equation}

\noindent where $\mathbf{u}_{m,n}$ denotes the unit vector directed
from the center of disk $n$ to the center of disk $m$,
\begin{equation*}
  \mathbf{u}_{m,n} = \frac{\mathbf{r}_m-\mathbf{r}_n}{\norm{\mathbf{r}_m-\mathbf{r}_n}},
\end{equation*}
and \noindent $\mathbf{u}_{m,w}=\pm\hat{\mathbf{y}}$ is a unit vector
directed into the channel from the wall that is nearer to disk~$m$.
To implement the periodic boundary condition, two extra disks ${N+1}$
and ${N+2}$, which are equivalent to disks $1$ and $2$, are added to
the end of the channel; this avoids having the matrix
$\boldsymbol{\mathsf{A}}$ depend on $L$, which would make the problem
nonlinear.  The bottom right section of $\boldsymbol{\mathsf{A}}$
containing these disks is then
\[
\bordermatrix{~ & \cdots & N-1,w & N-1,N & N-1,N+1 & N,w & N,N+1 & N,N+2 \cr
	      \vdots & \ddots &   &   &   &   &   \cr
	      {N-1} &  &  \mathbf{u}_{N-1,w} & \mathbf{u}_{N-1,N} & \mathbf{u}_{N-1,N+1} & 0 & 0 & 0 \cr
	      N &  & 0 &  -\mathbf{u}_{N-1,N} & 0 & \mathbf{u}_{N,w} & \mathbf{u}_{N,N+1} & \mathbf{u}_{N,N+2} \cr
	      {N+1} &  &  0 & 0 & -\mathbf{u}_{N-1,N+1} & 0 & -\mathbf{u}_{N,N+1} & 0 \cr
	      {N+2} &  &  0 & 0 & 0 & 0 & 0 & -\mathbf{u}_{N,N+2} \cr},
\]
\end{widetext}
and the displacements of disks ${N+1}$ and ${N+2}$ are defined by the
last two linear constraints, $\Delta \mathbf{r}_{N+1} = \Delta
\mathbf{r}_1$ and $\Delta \mathbf{r}_{N+2} = \Delta \mathbf{r}_2$.

The solution of the linear programming problem described above
provides a set of displacements that reduces the length of the system.
It is easily shown \cite{donev:04} that any set of displacements that
satisfy the {linearized} nonoverlap constraints will also satisfy the
exact (nonlinear) constraints.  Thus, to obtain the densest packing we
apply these displacements, recalculate $\boldsymbol{\mathsf{A}}$, and
then repeat the process until no further displacement of the disks
will reduce the length.  For our minimum-length configurations with
strict zigzag order, we find that there is no room for disks to
``rattle'': the system has reached a jammed state.  Our stopping
criterion was chosen as $\Delta L < 10^{-8}\sigma$, as this is a
typical observed order of magnitude for changes when the algorithm is
applied to a known stable configuration, such as the buckled crystal.
When a state is stable the final state is, of course, the same as the
initial state, to within the numerical precision of the calculation.

The sandwich states, with and without the insertion of extra pairs of
disks at positions $\pm y_1$, were compressed using the iterative
algorithm described above.  In the main text we have shown the result
of compressing the $N=24$ sandwich state containing no extra disks 
at~$\pm y_1$; see Fig.~3(c,d).  Another example for which the sandwich
state is found to be unstable is shown in
Fig.~\ref{fig:compressed_w_insertion} for the case $N = 32+10$, which
contains the maximum number of insertions of disks at~$\pm y_1$ into a
sandwich state of $32$ disks.  The configuration is unstable when only
one unit cell of buckled crystal is included (as shown in
Fig.~\ref{fig:compressed_w_insertion}(a)), but becomes stable if two
unit cells of buckled crystal are present.

\bibliography{refs}
\end{document}